\documentstyle[12pt,fleqn]{article}

\newcommand{\sectiona}[1]{\setcounter{equation}{0}\section{#1}
}

\newcommand{\A}{\alpha}
\newcommand{\B}{\beta}

\newcommand{\G}{\theta}

\newcommand{\X}{\xi}

\newcommand{\be}{\begin{eqnarray}}
\newcommand{\ee}{\end{eqnarray}}

\newcommand{\ket}[1]{| #1 \rangle}

\newcommand{\proj}[1]{| #1 \rangle \langle #1 |}
\newcommand{\inpr}[2]{\langle #1 | #2 \rangle}
\begin{document}

\title{Consistent histories and relativistic invariance in the modal interpretation of
quantum mechanics}

\author{Dennis Dieks\\ Institute for the History and Foundations of Science\\
Utrecht University, P.O.Box 80.000 \\ 3508 TA Utrecht, The
Netherlands}
\date{}
\maketitle

\begin{abstract}
Modal interpretations of quantum mechanics assign definite properties to physical
systems and specify single-time joint probabilities of these properties.
We show that a natural extension,
applying to properties at several times, can be given if a decoherence condition is
satisfied. This extension defines ``histories'' of modal properties. We
suggest a modification of the modal interpretation, that offers prospects of
a more general applicability of the histories concept.
Finally, we sketch a proposal to apply the procedure for finding histories and a
many-times
probability distribution to the context of algebraic quantum field
theory. We show that this leads to results that are relativistically invariant.
\end{abstract}
\vspace{3cm}
\newpage

\sectiona{Introduction}\label{sect1}
Modal interpretations of quantum mechanics \cite{vanfra,dieks1,healey,dieks2} 
interpret the mathematical formalism of quantum theory in terms of
properties possessed by
physical systems---in contradistinction to interpretations that take the formalism
as an instrument to calculate macroscopic measurement outcomes and their
probabilities.
By the statement that a system possesses a property we mean that some quantum
mechanical observable pertaining to the system takes on a definite value. However,
it is impossible to give all observables definite values while preserving the quantum
mechanical relations between them (as shown by the Kochen and Specker no-go
theorem).
Modal
interpretations therefore specify a {\em subset} of all observables; only the
observables in this subset are assigned definite values. It is characteristic of the
modal approach that this is done in a state-dependent way: the quantum mechanical
state of the system contains all information needed to determine the set of
definite-valued observables. 
The precise prescription for finding this set makes use of the Schmidt 
bi-orthogonal
decomposition of the composite state of a system plus its environment; or, more
generally, of the spectral
decomposition
of
the density operator describing a single system. 

Most work on the modal interpretation up to now has concentrated on properties
at a single time---with the exception of the pioneering studies of
Bacciagaluppi and Dickson \cite{baccia} and Vermaas \cite{vermaas2}.
No proposal for a joint probability distribution of physical
properties at several times has been generally accepted, and it has been queried
whether a
natural joint distribution can exist at all \cite{kent}. Moreover, the probability
distributions that have been proposed \cite{baccia} suffer from the problem that
they are not Lorentz-invariant; and it has been argued that it is a general feature of
modal dynamical schemes that they single out a preferred frame of
reference \cite{dickson}.

In this Letter we point out that in cases in which a decoherence condition is satisfied
there is a natural and obvious
candidate for a joint many-times probability distribution (in spite of the arguments to
the
contrary in \cite{kent}). We attempt to extend the applicability of this joint 
probability distribution by suggesting a modification of the modal interpretation
scheme. Finally, we consider the application of the ``histories'' idea to quantum field
theory (assuming decoherence again). Under certain conditions, the implementation of the idea here leads to
a Lorentz-invariant joint probability distribution of properties associated with very small
regions (approximating
space-time points). 

\sectiona{The modal scheme} 

Let $\A$ be
our system and let $\B$
represent its total environment (the rest of the universe).
Let $\A\&\B$  be
represented by
$\ket{\psi^{\A\B}} \in {\cal H}^{\A} \otimes {\cal H}^{\B}$. 
The
bi-orthonormal decomposition of $\ket{\psi^{\A\B}}$,
\be
\ket{\psi^{\A\B}} & = & \sum_{i} c_{i} \: \ket{\psi^{\A}_{i}}
\:
\ket{\psi^{\B}_{i}} \label{eq1} \;\;\; ,
\ee
with $\inpr{\psi^{\A}_{i}}{\psi^{\A}_{j}} =
\inpr{\psi^{\B}_{i}}{\psi^{\B}_{j}} = \delta_{ij}$, generates
a set of projectors operating on ${\cal H}^{\A}$: 
$\{\proj{\psi^{\A}_{i}} \}_{i}$.
If there is no degeneracy among the numbers $\{ | c_{i} |^{2}
\}$ this is a uniquely determined set of one-dimensional projectors.
If there is degeneracy, the projectors
belonging to one value of
$\{ | c_{i} |^{2} \}$ can be added to form a
multi-dimensional projector; the thus generated new set of
projectors, including multi-dimensional ones, is again uniquely determined. These
projectors are the ones occurring in the spectral
decomposition of
the reduced density operator of $\A$.  

The modal interpretation assigns definite values to the
subset of all physical magnitudes that is generated by these projectors; i.e., the subset
obtained by
starting with these projectors, and then including their continuous functions, real
linear combinations, and
symmetric and antisymmetric products \cite{clifton}
(the thus defined real, closed in norm, linear subspace of all observables constitutes
the set of ``well-defined'' or
``applicable'' physical magnitudes, in Bohrian parlance). 
{\em Which}
value among the possible values of a definite magnitude
is actually realised is not fixed by the interpretation. For
each possible value a probability
is specified: the
probability that the magnitude represented by
$\proj{\psi^{\A}_{i}}$ has
the value
$1$ is given by
$|c_{i}|^{2}$. In the case of degeneracy it is stipulated that
the magnitude
represented by $\sum_{i\in I_{l}}
\proj{\psi^{\A}_{i}}$ has value $1$ with probability
$\sum_{i\in I_{l}} |c_{i}|^2$ ($I_{l}$ is the index-set
containing indices $j$, $k$ such that $|c_{j}|^{2} =
|c_{k}|^{2}$).

The observation that the definite-valued projections occur in the spectral
decomposition of $\A$'s density operator gives
rise
to a generalisation of the above scheme that is also applicable to the case in
which the total system $\A\&\B$
is not represented by a pure
state: find $\A$'s density operator by partial tracing from the total density operator,
determine its spectral resolution
and construct the set of definite-valued observables from the projection operators in
this spectral resolution
\cite{verm}.

The above recipe for assigning properties is meant to apply to each physical system in
a non-overlapping collection of
systems that together make up the total universe \cite{baccia,dieks3}. It is easy to
write down a satisfactory {\em joint}
probability distribution for the properties of such a collection (or a subset of it):
\begin{equation}
Prob(P^{\A}_{i}, P^{\B}_{j},...., P^{\G}_{k},...., P^{\X}_{l})
= \langle \Psi | P^{\A}_{i}.
P^{\B}_{j}.....P^{\G}_{k}.....P^{\X}_{l} | \Psi
\rangle ,\label{prob}
\end{equation}
where the left-hand side represents the joint probability for the
projectors occurring in the argument of taking the value $1$, and where $\Psi$ is the
state of the total system consisting of
$\A$, $\B$, $\G$, etc. \cite{verm}.
It is important for the consistency of this probability
ascription that the projection operators occurring in the
formula all commute (which they do, since they
operate in different non-overlapping Hilbert spaces).

However, a full-fledged interpretation of quantum mechanics in terms of properties
of physical systems should not only
specify the
class of definite-valued observables at each instant, and their probability distribution,
but should also make it clear
what the probability is that a value present at one instant of time goes over
into a given possible value at another time; or, equivalently, what the joint probability
is of
values of definite-valued observables
at several times. It turns out that it is not easy to find a compelling and natural 
solution to this problem for the general case
\cite{baccia,vermaas2}. But in the special, though physically important, case that a
decoherence
condition is satisfied there
{\em is} such a natural joint probability for histories of properties, as we will now
discuss.
 
\sectiona{Consistent histories of modal properties}

There is a natural analogue of expression (\ref{prob}) for the case of Heisenberg
projection operators pertaining to
different instants of time:
\begin{eqnarray}
\lefteqn{Prob(P_{i}(t_{1}), P_{j}(t_{2}),.... P_{l}(t_{n}))
=} \nonumber \\ & & \langle \Psi | P_{i}(t_{1}).
P_{j}(t_{2}).....P_{l}(t_{n}) . P_{l}(t_{n})
. .... P_{j}(t_{2}) .  P_{i}(t_{1}) | \Psi \rangle .\label{sevtimeprob}
\end{eqnarray}
This expression is in accordance with the standard, ``Copenhagen'', prescription for
calculating the joint probability
of outcomes of consecutive measurements. It agrees also with the joint distribution
assigned to ``consistent histories'' in the consistent histories approach to the
interpretation of quantum mechanics \cite{grif}. However, it should be noted that the
projection
operators in (\ref{sevtimeprob}),
pertaining to different times as they do, need not commute. As a result,
(\ref{sevtimeprob})
does not automatically yield a consistent probability distribution. For this reason it is
an
essential
part of the consistent histories approach to impose the
following decoherence condition, in
order to guarantee that expression (\ref{sevtimeprob}) is an ordinary
Kolmogorov probability:
\begin{eqnarray}
\lefteqn{\langle \Psi | P_{i}(t_{1}).
P_{j}(t_{2}).....P_{l}(t_{n}) .
P_{l^{\prime}}(t_{n}) ......
P_{j^{\prime}}(t_{2}) .  P_{i^{\prime}}(t_{1}) | \Psi \rangle =0}
\nonumber \\ & & i \neq
i^{\prime} \vee j \neq j^{\prime} \vee
....\vee l \neq l^{\prime}.\label{consist}
\end{eqnarray}

In the consistent histories approach all sequences of properties which satisfy
the decoherence condition (\ref{consist}) are considered. A problem of this approach
is that the decoherence condition leaves room for many, mutually incompatible,
families of consistent properties, so that the set of definite-valued properties is not
fully determined. By contrast, the modal interpretation provides an unequivocal rule
to fix the definite-valued properties. It seems therefore worth-while to investigate
whether the two approaches can be combined by using the above probability
distribution for modal histories. 
However, in Ref.\ \cite{kent} it is argued that the projection operators singled out 
as definite-valued by the
modal interpretation will not
satisfy the decoherence condition (except in the very special and
physically
unrealistic circumstance in which the system's properties evolve deterministically). If
valid, this argument would
make the probabilistic resources of the consistent histories approach unavailable
to the modal interpretation. 

The essential ingredient of the argument is the following (see the text
accompanying Eq.\ (2.6) in Ref.\ \cite{kent}). Consider the decoherence condition
pertaining to two instants, $t_{1}$ and $t_{2}$, and let $P_{l}(t_{2}) =
\proj{\psi_{l}(t_{2})}$.
We then have
$\langle \Psi | P_{i}(t_{1}) | \psi_{l}(t_{2}) \rangle \langle \psi_{l}(t_{2}) |
P_{i^{\prime}}(t_{1}) | \Psi \rangle = 0 $,
if $i \neq i^{\prime}$, for all values of $l$. That means, the argument goes, that
either
$ \langle \Psi | P_{i}(t_{1}) | \psi_{l}(t_{2}) \rangle = 0 $ or
$ \langle \psi_{l}(t_{2}) | P_{i^{\prime}}(t_{1}) | \Psi \rangle = 0 $. From this it
follows that
$\ket{\psi_{l}(t_{2})}$
is orthogonal to all but one of the states $\ket{ P_{i}(t_{1}) \Psi}$. That would
imply a deterministic evolution of
properties, and if the projectors  $P_{i}(t_{1})$ are one-dimensional the properties at
time
$t_{2}$ would even have to be the same as those at time $t_{1}$.

The questionable premise in this argument is 
the presupposition that
$ \langle \Psi | P_{i}(t_{1}) | \psi_{l}(t_{2}) \rangle$ and
$ \langle \psi_{l}(t_{2}) | P_{i^{\prime}}(t_{1}) | \Psi \rangle$ are numbers,
instead
of a bra and a ket, respectively.
Making the assumption that these expressions represent numbers is equivalent to
assuming that
$\ket{\Psi}$ is an element of the Hilbert space
of the system under consideration.  This assumption does not fit in with the
modal approach: the modal property
ascription, as explained above, uses for $\ket{\Psi}$ the state of the
combination ``system and rest of the universe''. As we shall show in a moment,
the fact that the presence of an environment has always to be taken into
account not only invalidates the above argument, but also makes it natural and easy
to incorporate the idea of
decoherence in the modal scheme so that condition (\ref{consist})
is satisfied. As a consequence, Eq.\ (\ref{sevtimeprob}) yields
a
consistent
joint multi-times
probability distribution for modal properties in the case in which this decoherence
condition is fulfilled.

The notion of decoherence to be used is the following. It is a general feature of the
modal interpretation that if a system acquires a
certain property, this happens by virtue of its interaction with the environment,
as expressed in Eq.\ (\ref{eq1}). As can be seen from this equation, in this process
the system's
property
becomes correlated with a property of the environment. Decoherence is now defined
to
imply the irreversibility of this process of correlation formation: the rest of the
universe retains a trace of the
system's property, also at later times when the properties of the system itself may
have
changed.
In other words, the rest of the universe acts as a memory of the properties the system
has had; and decoherence guarantees that this memory remains intact. For the state
$\ket{\Psi}$ this means
that in the Schr\"{o}dinger picture it can be written in the following form: 
\be
\ket{\Psi(t_{n})} & = & \sum_{i,j,...,l} c_{i,j,...,l} \: \ket{\psi_{i,j,...,l}}
\:
\ket{\Phi_{i,j,...,l}} \label{eqdeco} \;\;\; ,
\ee
where $\ket{\psi_{i,j,...,l}}$ is defined in the Hilbert space of the system,
$\ket{\Phi_{i,j,...,l}}$
in the Hilbert space pertaining to the rest of the universe, and where
$\langle \Phi_{i,j,...,l} | \Phi_{i^{\prime},j^{\prime},...,l^{\prime}} \rangle =
\delta_{i i^{\prime} j j^{\prime}...l l^{\prime}}$. 
In (\ref{eqdeco}) $l$ refers to
the properties $P_{l}(t_{n})$, $j$ to the properties $P_{j}(t_{2})$, $i$ to the
properties
$P_{i}(t_{1})$, and so on. 

The physical picture that motivates a $\ket{\Psi}$ of this form is that
the final state results from
consecutive measurement-like interactions, each of which is responsible for generating
new properties. Suppose that in the first interaction
with the environment the properties $ \proj{\A_{i}}$ become definite: then the state
obtains the form $\sum_{i} c_{i} \ket{\A_{i}} \ket{E_{i}}$, with $\ket{E_{i}}$
mutually
orthogonal states of the environment. In a subsequent interaction, in which the
properties $ \proj{\B_{j}}$ become definite, and in which the environment
``remembers'' the presence of the $\ket{\A_{i}}$, the state is transformed into 
$\sum_{i,j} c_{i} \inpr{\B_{j}}{\A_{i}} \ket{\B_{j}} \ket{E_{i,j}}$, with mutually
orthogonal environment states $\ket{E_{i,j}}$. Continuation of this series of
interactions eventually leads to Eq.\ (\ref{eqdeco}), with in this case $
\ket{\psi_{i,j,...,l}} =
\ket{\psi_{l}}$ (see below for a generalisation).   

If this picture of consecutive measurement-like interactions applies, it follows that
in the Heisenberg picture we have
$P_{l}(t_{n}) ......
P_{j}(t_{2}) .  P_{i}(t_{1}) \ket{\Psi} = c_{i,j,...,l} \ket{\psi_{i,j,...,l}}
\ket{\Phi_{i,j,...,l}}$. Substituting this in the expression at the left-hand side of Eq.\
(\ref{consist}), and making use of the orthogonality properties of the states
$\ket{\Phi_{i,j,...l}}$, we find immediately that the consistent histories
decoherence
condition (\ref{consist}) is satisfied. As a result, expression (\ref{sevtimeprob})
yields a classical
Kolmogorov
probability distribution of the modal properties at several times.

\sectiona{A possible modification of the modal scheme}

The just-discussed decoherence scenario is based on the assumption that in
consecutive
interactions the initial state of the system is not relevant for the properties that are
generated:
we wrote that $\ket{\A_{i}} \ket{E_{i}}$ is transformed into 
$\sum_{j} \inpr{\B_{j}}{\A_{i}} \ket{\B_{j}} \ket{E_{i,j}}$, with the same set
$\{\ket{\B_{j}}\}$
for all values of $i$. In some physically important cases this is not 
unrealistic. The prime example is the case in which the object is subjected
to consecutive measurements by devices designed to measure certain observables,
regardless of the
state of the incoming system. A similar situation can occur in cases in which no
instruments
constructed by humans are present. For instance, interactions by means of a 
potential that depends only on the object observable $A$, with an environment with
very many degrees of freedom,
will tend to destroy the coherence between object states
that correspond to different $A$-values and can thus be regarded as
(approximate) $A$-measurements, regardless of the object's initial state. Consecutive
exposures to such environments
can be described by (\ref{eqdeco}).

But in general we will have to consider the situation in which no such
measurement-like interactions occur,
and in which the type of the later interaction may depend 
on the result of an earlier interaction. In this general case we
can represent what happens in an interaction by writing:
\be
\sum_{i} c_{i} \ket{\A_{i}} \ket{E_{i}} \longrightarrow
\sum_{i,j} c_{i,j} \inpr{\B_{i,j}}{\A_{i}} \ket{\B_{i,j}} \ket{E_{i,j}},\label{evol}
\ee
with $\inpr{\B_{i,j}}{\B_{i,j^{\prime}}}= \delta_{jj^{\prime}}$. In this formula a 
biorthogonal decomposition has been
written down for each value of $i$ separately; in other words, we are looking at the
various mutually orthogonal ``branches'', that result from an interaction, separately
and look how each of them branches itself in a subsequent interaction. A series of
consecutive interactions described in this way again leads to
a total final state of the form (\ref{eqdeco}). However, the biorthogonal
decomposition
applied to the total right-hand side of (\ref{evol}) would now in general
not yield the projectors
$\proj{\B_{i,j}}$ as definite-valued observables. Therefore,
$P_{j}(t_{2}) .  P_{i}(t_{1}) \ket{\Psi}$, with $P(t_{1,2})$ the projectors
that {\em are} singled out by the biorthogonal decomposition as
definite-valued at $t_{1}$ and $t_{2}$, respectively, will not be equal to
$ c_{i,j} \inpr{\B_{i,j}}{\A_{i}} \ket{\B_{i,j}} \ket{E_{i,j}}$ (the projectors
$P_{j}(t_{2})$ will
generally not commute with the projectors $\proj{\B_{i,j}}$). This has the
consequence that there is no reason to
expect that the decoherence condition (\ref{consist}) will be satified. 

This suggests a modification of the modal scheme, designed to guarantee that Eq.\
(\ref{sevtimeprob}) remains
valid as the joint probability of possessed properties, even in the more general
situation just described.
The idea is to change the property assignment in such a way that the right-hand side
of Eq.\ (\ref{evol})
comes to represent a situation in which the projectors $\proj{\B_{i,j}}$ {\em are}
definite-valued. Such a property ascription would be in accordance with the intuition
that the different branches are irrelevant to each other as long as no 
re-interference occurs. As in the many-worlds interpretation, in which we do not need
to consider what happens in other worlds in order to describe what happens in our
world, it is proposed here that the ascription of properties, as a result of
interactions, should be done {\em per branch}.

To this end, we may posit as an interpretational rule that in a composite 
state of the form
\be
\sum_{i,j} c_{i,j} \inpr{\B_{i,j}}{\A_{i}} \ket{\B_{i,j}} \ket{E_{i,j}},
\label{newmodal}
\ee
with $\inpr{E_{i,j}}{E_{i^{\prime},j^{\prime}}} = \delta_{i i^{\prime} j
j^{\prime}}$, and
$\inpr{\B_{i,j}}{\B_{i,j^{\prime}}}= \delta_{jj^{\prime}}$, resulting from an
interaction of the form
$\sum_{i} c_{i} \ket{\A_{i}} \ket{E_{i}} \longrightarrow
\sum_{i,j} c_{i,j} \inpr{\B_{i,j}}{\A_{i}} \ket{\B_{i,j}} \ket{E_{i,j}}$, 
the projectors $\proj{E_{i,j}}$ represent properties of the environment, 
and that 
the system has properties $\proj{\B_{i,j}}$, in one-to-one correlation to these
environmental properties. In terms of the biorthogonal decomposition, the new
proposal says that
the system's properties are determined by the separate
decomposition that can be written down for each value of $i$: the 
terms of a biorthogonal decomposition that do not recombine (interfere) in
subsequent interactions are treated as individual branches, isolated from the other
terms. The various branches are assigned definite-valued observables through their
own individual
biorthogonal decomposition.

It should be noted that the state $\ket{\Psi}$ of the total system, together with the
total Hamiltonian, uniquely determine the properties ascribed in this scheme. This is
because the Hamiltonian governs the evolution, so that the initial state---before
interactions started---is fixed by $\ket{\Psi}$ and the Hamiltonian. Further, the
branching that occurs in the subsequent interactions is fully determined by writing
down the biorthogonal decomposition (per branch) after each interaction.
 
According to this way of interpreting the quantum mechanical state, the
properties $P_{i}(t_{n})$ that
are assigned satisfy Eq.\ (\ref{consist}) and make Eq.\ (\ref{sevtimeprob}) a
consistent probability distribution (if there is decoherence, that is, as long as there is
no interference of the several branches).
This brings the modal scheme closer to the consistent histories
interpretation, in which exactly those histories are considered for which
Eq.\ (\ref{sevtimeprob}) yields a consistent probability distribution. There would still
be a
distinction, though: the requirement that
Eq.\ (\ref{sevtimeprob}) be a consistent probability is by itself not enough
to determine what properties are definite at the various instants---in other words,
what
the family of consistent histories is. There can be many mutually inconsistent
but partially overlapping families of consistent histories \cite{dowker}). This
constitutes
a problem in the
consistent histories approach, because the consistency condition is the only constraint
one
has in that approach. By contrast, in our suggested modified modal scheme we retain
the distinctive
modal feature that
the definite-valued observables are uniquely defined by a fixed rule. As a
consequence, there is only one family
of consistent
histories to consider.

We do not further discuss and elaborate this suggested modification of the modal scheme
here, but instead turn to the question of whether the notion of histories of definite
properties,
with well-defined probabilities, can also be applied outside of the context
of non-relativistic quantum theory, to relativistic quantum field theory.
It seems clear that any interpretation of quantum theory should at least have the prospect of being so applicable in
order to be taken seriously.

\sectiona{Application to relativistic quantum field theory}

The traditional interpretational problems of quantum mechanics exist no less in
quantum field theory; and as we will see shortly there are also additional problems.
Because quantum field theory occupies a central place in the present physical description of the world,
any interpretation of quantum theory should at least offer prospects of leading to sensible
results if applied to this new context. We will therefore now briefly outline a method to implement the modal ideas
of the foregoing to quantum field theory.

As before,
the central issue is that it is not obvious that the theory is about objective physical
states of affairs, even
in circumstances in which no macroscopic measurements are being made. The  fields
in quantum field theory do not attach values
of physical magnitudes to space-time points. Rather, they are fields of operators, with
a standard interpretation
in terms of macroscopic measurement results. 
In accordance with what we have said about the interpretation of the
Schr\"{o}dinger-Heisenberg theory we
would like to give another meaning to the formalism, namely in terms of
physical systems that possess certain properties.
We will take as our framework the formalism of local algebraic quantum field
theory as explained
in \cite{haag}, because of its generality. In this framework, a C$^{\ast}$-algebra of observables is associated
to each open region
of Minkowski space-time. What we would like to
do is to provide
an interpretation in which not only operators, but also {\em
properties} are assigned to
space-time regions. That is, we would like at least some of the observables to have
definite values. This would lead
to a picture in which it is possible to speak of objective {\em
events} (if some physical
magnitude takes on a definite value in a certain spacetime region, this constitutes the
event that this magnitude has that value there and then).

If the open space-time regions could be regarded as physical systems, each one of
them described in a factor space of a total Hilbert space (this total Hilbert space
would then be the tensor product of the spaces belonging to non-overlapping subsystems),
an immediate generalisation
of the modal scheme would be possible. However, algebraic quantum field theory is much
less amicable to the notion of a localised physical system than might be expected. The
local algebras are of type III, and this implies that they cannot be represented as
algebras of bounded observables on a Hilbert space (such algebras are of type I). 
We can therefore not take the open space-time regions and their algebras as
fundamental, if we want an interpretation in terms of (more or less) localised systems whose properties
would specify an event. Such an interpretation is highly desirable, however, at least in the limiting situation in which
classical concepts become applicable: it should be possible, in this limiting situation, to speak of field values
in small space-time regions.

One possible way out is to use the algebras of type I that ``lie
between two local algebras''. That such type-I algebras exist is assumed in the
postulate of the ``split property'' (\cite{haag}, Ch.\ V.5, \cite{summers}). We accept this postulate and
focus on the algebras of type I lying between two concentric standard ``diamond'' regions
with radii $r$ and $r+\epsilon$,
respectively, with $r$ and $\epsilon$ small numbers. Two of such type-I algebras, defined
with respect to two double diamonds associated with regions 
that have space-like separation, are independent in a strong sense: the total algebra generated by them
can be represented by the tensor product of the two algebras, defined on the tensor product of two representative
Hilbert spaces  $\cal{H}_{A}$ and $\cal{H}_{B}$ of the two individual algebras separately: $\cal{H}_{A \& B}=\cal{H}_{A}
\otimes \cal{H}_{B}$ \cite{summers}.

In this way we may
justify the approximate validity of the notion of a small space-time region
as a physical subsystem represented in
a factor space of the total Hilbert space. Of course, there is arbitrariness here, for example in choosing the positions
of the double
diamonds in the manifold and in fixing the
values of $r$ and $\epsilon$; it is not to be expected that a sensible algebra will result if  $r\rightarrow 0$.
This reflects the fact that the structure of quantum field theory, despite
first appearances, by itself does not provide a natural arena for a space-time picture in which physical magnitudes attach to
small space-time regions. We expect therefore that a space-time interpretation,
like the one we are trying to construct, will not have a fundamental status but can only be employed
in a classical limiting situation in which classical field and particle concepts become approximately
applicable. In other words, we do not assume that actually field values {\em are}
associated with space-time points, and that our description provides an approximation to that real state of affairs.
Rather, we think that in
the general case the ordinary space-time picture will not be possible and that the classical field and particle
concepts can only
be applied in a limiting situation; accordingly, in the general case $x$ and $t$ will be parameters that do not possess their
usual space-time meaning. The approximate applicability of classical concepts in a limiting situation
is perhaps connected to the presence of 
decoherence mechanisms that
decouple certain $x$, $t$ regions from each other (see also below; we will need such an assumption of
decoherence anyway, to arrive at consistent joint probabilities).           

We will now adopt the modal ideas of section 3 to assign values to the
observables in the just-defined type-I algebras, loosely associated with small space-time regions.
We will therefore assume that there is
decoherence in the
sense discussed in section 3. This assumption is natural in the context of local
quantum physics \cite{schroer}, because
the physical systems in local
quantum physics
automatically have an infinite environment to interact
with, so that decoherence and irreversibility can easily occur.

The core idea of the modal interpretation is to select a subset of definite-valued
observables from the algebra
of all observables by means of an objective,
fixed, rule. This was implemented within the framework of non-relativistic
quantum mechanics via the selection principle
based on the biorthogonal decomposition of the total state or, more generally, the
spectral resolution
of the system's density operator.
Here, we will analogously consider the projectors occurring in
the spectral resolution of the
density operator (defined in a factor space) representing the partial state defined on the type-I algebra of
observables
attached to the very small ``split-regions'' that approximate space-time points. At each instant, we will consider
a collection of such regions that have space-like separation with respect to each other, and can
be described by means of the tensor product of the individual factor spaces.
The selected projectors provide us with a
base set of definite-valued
quantities. As before, the complete collection of definite-valued observables can be
constructed
from this base set by closing the set under the operations of taking continuous
functions, real linear combinations,
and symmetric and antisymmetric products \cite{clifton}. The
probability of projector
$P_{l}$ having the value $1$ is $\langle \Psi | P_{l} | \Psi \rangle$. Subdividing (approximately) the whole
of Minkowski space-time into a collection of non-overlapping point-like regions, that have
space-like separation on each simultaneity hyperplane, and
applying the above prescription to the associated algebras, we achieve the picture
aimed
at: to each (approximate) space-time point belong
definite values of
some physical magnitudes, and this constitutes an event at the position and time in
question.

In order to complete this picture we should specify the joint probability of events
taking place at different space-time
``points''. It is natural to consider, for this purpose, a generalisation of expression
(\ref{sevtimeprob}). The first problem
encountered in generalizing this expression to the relativistic context is that we no
longer have absolute time available
to order the sequence $P_{i}(t_{1})$, $P_{j}(t_{2})$, ...., $P_{l}(t_{n})$.  In
Minkovski space-time we only have the
partial ordering $y < x$ (i.e., $y$ is in the causal past of $x$) as an objective relation
between space-time points. However, we can still impose a linear ordering on the
space-time points in any region in
space-time by considering equivalence classes of points which all have space-like
separation with respect to each other---for instance, points whose centres are on the same simultaneity hyperplane.
Of course, there are infinitely many
ways of subdividing the region into such
space-like collections of points. It will have to be shown that the joint probability
distribution that we are going to
construct is independent of the particular subdivision that is chosen.

Take one particular linear time ordering of the points in a closed region of
Minkowski space-time, for instance one generated
by a set of parallel simultaneity hyperplanes (i.e.\ hyperplanes that are all Minkowski-orthogonal
to one given time-like worldline).
Let the time parameter $t$ label very thin slices of space-time (approximating
hyperplanes) in which ``points'' of the kind introduced above, with mutual space-like
separation, are located. We can now write
down a joint probability distribution
for the properties, in exactly the same form as in
Eq.\ (\ref{sevtimeprob}): 
\begin{eqnarray}
\lefteqn{Prob(P^{\ast}_{i}(t_{1}), P^{\ast}_{j}(t_{2}),.... 
P^{\ast}_{l}(t_{n}))=} 
\nonumber \\ & & \langle \Psi | P^{\ast}_{i}(t_{1}).
P^{\ast}_{j}(t_{2}).....P^{\ast}_{l}(t_{n}) . P^{\ast}_{l}(t_{n})
. .... P^{\ast}_{j}(t_{2}) .  P^{\ast}_{i}(t_{1}) | \Psi \rangle.
\label{sevtimeprob1}
\end{eqnarray} 
In this formula the projector $P^{\ast}_{m}(t_{l})$ represents the
properties of the space-time ``points'' on the ``hyperplane'' labeled by $t_{l}$.
That is: 
\begin{equation}
P^{\ast}_{m}(t_{l}) = \Pi_{i} P_{m_{i}}(x_{i}, t_{l}), 
\label{product}
\end{equation}
with $\{ x_{i} \}$ the central positions of the point-like regions considered on the
``hyperplane''.
The index $m$ is symbolic for the set of indices $\{m_{i}\}$.
As explained above, the modal proposal is that the projectors
$P_{m_{i}}(x_{i},t_{l})$ come
from the spectral decomposition of the mixed state defined on the algebra 
belonging to the space-time ``point'' $(x_{i},t_{l})$ (we do not pursue here the
change
in this if
we
consider the modification suggested in section 4). Because all the considered 
point-like regions on the ``hyperplane'' $t_{l}$ are space-like separated from each
other,
the associated projectors commute (the principle of micro-causality). This important
feature of local quantum physics 
guarantees that the product operator of Eq.\ 
(\ref{product}) is again a projection operator, so that expression 
(\ref{sevtimeprob1}) can be treated in the same way as Eq.\ (\ref{sevtimeprob}).
In particular, we will need an additional condition to ensure that
(\ref{sevtimeprob1})
yields a Kolmogorovian probability.

The decoherence condition that we are going to use is essentially the same as discussed
in
section 3. Suppose that at
space-time ``point'' $(x,t)$ the magnitude represented by the set of projector
operators $\{P_{k}\}$ is definite-valued; $P_{l}$ has value $1$, say. 
Intuitively speaking, the notion of decoherence that we invoke is 
that in the course of the further evolution there subsists
a trace of $P_{l}(x,t)$ in the future lightcone of 
$(x,t)$. 
That is, decoherence implies that on each space-like hyperplane 
intersecting the future lightcone of $(x,t)$ there is at least one space-time 
point, within this lightcone, that has a property strictly correlated to
$P_{l}$. One way of 
fulfilling this 
decoherence 
condition would be given by the existence of a time-like worldline (a propagating signal) 
going through
$(x,t)$ (or a bundle of such worldlines), each point of which is characterized 
by the value $1$ of $P_{l}(x,t)$ (apart from evolution). 
This makes explicit the idea
that the effects of the irreversible interaction responsible for decoherence 
propagate within the future lightcone of $(x,t)$. 

If this decoherence condition is fulfilled, we have that on the hyperplane
$t_{2}$ at least one of the space-time points, say $(x^{\prime},t_{2})$,
has properties $\{P^{\prime}_{k}\}$  correlated to $\{P_{k}\}$: $P^{\prime}_{k}.P_{k^{\prime}}| \Psi \rangle = 0$
if $k \neq k^{\prime}$.

In that case we not only have that 
that $\langle \Psi | P_{l}(x,t_{1}) P^{\prime}_{k}(x^{\prime},t_{2}) 
P^{\prime}_{k^{\prime}}(x^{\prime},t_{2}) P_{l}(x,t_{1}) | \Psi \rangle = 0 $ if
$k \neq k^{\prime}$ (this follows simply from the orthogonality of $P_{k}$ and
$P_{k^{\prime}}$, expressing the incompatibility of two 
different values of the same observable at one space-time point), but we also find 
$\langle \Psi | P_{l}(x,t_{1}) P^{\prime}_{k}(x^{\prime},t_{2}) 
P^{\prime}_{k^{\prime}}(x^{\prime},t_{2}) P_{l^{\prime}}(x,t_{1}) | \Psi \rangle = 0$
if $l \neq l^{\prime}$, both if $k = k^{\prime}$ and if $k \neq k^{\prime}$.
This expresses that possibilities
characterized by different values of an observable remain incompatible 
in the course of time. 
Assuming decoherence for the properties of all space-time points,
we find the analogue
of (\ref{consist}):
\begin{eqnarray}
\lefteqn{\langle \Psi | P^{\ast}_{i}(t_{1}).
P^{\ast}_{j}(t_{2}).....P^{\ast}_{l}(t_{n}) .
P^{\ast}_{l^{\prime}}(t_{n}) ......
P^{\ast}_{j^{\prime}}(t_{2}) .  P^{\ast}_{i^{\prime}}(t_{1}) | \Psi \rangle
=0}
\nonumber \\ & & i \neq
i^{\prime} \vee j \neq j^{\prime} \vee
....\vee l \neq l^{\prime}.\label{consist1}
\end{eqnarray}
This makes (\ref{sevtimeprob1}) a consistent Kolmogorovian 
joint probability for the joint occurrence of the events represented by
$P^{\ast}_{i}(t_{1})$, $P^{\ast}_{j}(t_{2})$, ....$P^{\ast}_{l}(t_{n})$.

The projectors $P^{\ast}_{i}(t_{1})$, $P^{\ast}_{j}(t_{2})$,
....$P^{\ast}_{l}(t_{n})$ depend for their definition on the chosen
set of hyperplanes, labeled by $t$. Therefore
(\ref{sevtimeprob1})
is not manifestly Lorentz invariant. However, the projectors $P^{\ast}(t)$ are
products of
projectors pertaining
to the individual space-time points lying on the $t$-hyperplanes, so
(\ref{sevtimeprob1}) can
alternatively be written in terms of these latter projectors. The specification
of the
joint probability of the physical quantities at all considered ``points'' in a given
space-time region
requires
(\ref{sevtimeprob1}) with all individual projectors appearing in it. Depending on the way in which the space-time region
has been
subdivided in space-like hyperplanes
in the definition of $P^{\ast}_{i}(t_{1})$, $P^{\ast}_{j}(t_{2})$,
....$P^{\ast}_{l}(t_{n})$, the individual projectors occur in
different orders in this complete
probability specification. However, there is a lot of conventionality in this ordering.
All operators attached to
point-like regions with space-like separation commute, so that their ordering can be
arbitrarily
changed. The only 
characteristic
of the ordering that is invariant under all these allowed permutations is that if $y <
x$ (i.e.\ $y$ is in the
causal past of $x$), $P(y)$ should appear before $P(x)$ in the expression for the joint
probability. But this
is exactly the characteristic that is common to all expressions that follow from writing
out (\ref{sevtimeprob1}),
starting from all different ways of ordering events with a time parameter $t$. All
these expressions can therefore
be transformed into each other by permutations of projectors belonging to space-time
``points'' with space-like separation.
The joint probability thus depends only on how the events in the space-time region
are ordered with respect
to the Lorentz-invariant relation $<$; it is therefore Lorentz-invariant itself.   
 
\sectiona{Conclusion}

The main aim of this Letter has been to show that a decoherence condition, analogous to the one whose fulfilment
is assumed in the consistent
histories approach,
can be satisfied in the modal interpretation of quantum mechanics. If the condition is
satisfied, a simple
and natural joint probability can be specified for the values of definite-valued
observables at several times.
This probability expression is the same as the one proposed in the consistent histories
scheme, or in traditional
quantum measurement theory. It is here combined with the characteristic feature
of the modal interpretation,
namely that the set of definite-valued observables at each instant is determined by an
objective rule that uses
only the form of the quantum mechanical state; this rule selects one family of
consistent histories. 

Because quantum field theory is central in present-day theoretical physics,
we have proposed a way of applying the modal histories scheme to that theory.
This proposal was motivated by the desire to obtain, at least in the classical limiting situation,
a picture in which events occur, in small space-time regions approximating space-time points. In this way it
becomes possible in principle to recover the classical picture according to which field values are attached to
space-time points.
We have
argued that the assumption of irreversible decoherence, combined with the modal
ideas, indeed offers prospects of obtaining
a Lorent-invariant account in which field magnitudes take on definite values in 
point-like regions and in which a
consistent joint probability for such values can be defined.    

\section*{Acknowledgement}

I am indebted to Dr.\ Klaas Landsman and Prof.\ Miklos Redei for valuable help concerning questions
in quantum field theory.

\end{document}